\documentstyle[aps,prd]{revtex}
\begin{document}
\newcommand{\be}{\begin{equation}}
\newcommand{\ee}{\end{equation}}
\newcommand{\ben}{\begin{eqnarray}}
\newcommand{\een}{\end{eqnarray}}
\newcommand{\nn}{\nonumber}
\newcommand{\n}{\label}

\title{Cosmic acceleration without quintessence}

\author{
Narayan Banerjee\footnote{Permanent address: Relativity and Cosmology
Research Centre, Department of Physics, Jadavpur University, 
Calcutta-700032, India} and 
Diego Pav\'{o}n\footnote{Electronic mail address: diego@ulises.uab.es}}
\address{Departamento de F\'{\i}sica. Facultad de Ciencias.
Edificio Cc. Universidad Aut\'onoma de Barcelona. E--08193 Bellaterra
(Barcelona). Spain}

\date{\empty}

\maketitle


\begin{abstract}
It is argued that the Brans--Dicke theory may explain the present 
accelerated expansion of the universe without resorting
to a cosmological constant or quintessence matter. 
\end{abstract}

\section{Introduction}
For several decades the general belief in cosmology had been
in favour of a  presently matter--dominated universe, expanding 
as  $a \propto t^{2/3}$, where $a$ is the scale factor of the
Robertson--Walker metric and $t$ is the cosmic time. The corresponding 
decelerated expansion ($ q_{0} =- a \ddot{a}/\dot{a}^2 = 1/2$) 
was more or less  compatible with all the cosmological tests. 
The problems with the standard cosmology were confined
to the early stages of the evolution of
the universe and they were expected to be taken care of by the
inflationary paradigm. But the recent observations regarding the
luminosity--redshift relation of the type Ia supernovae up to about
$z=1$ revealed that the universe is in fact expanding at a faster
rate, even possibly with an acceleration ($q_{0} < 0$) \cite{Perlmutter}. 
This observation naturally leads to the search for the matter fields,
hitherto unknown or neglected, which could introduce such a
non-decelerated expansion. This matter is called a quintessence
matter ( Q-matter for short). The list of possible candidates, being 
explored as Q-matter, consists of a cosmological constant or a time dependent
cosmological term, a scalar field with a potential giving
rise to a negative pressure \cite{Caldwell} at the present epoch, a dissipative
fluid yielding an effective negative stress \cite{Chimento} or 
more exotic matter like a frustrated network of non-Abelian cosmic 
strings or a frustrated network of domain walls \cite{Bucher}. 
Amongst the scalar fields chosen as the Q-matter, the ``tracker" field 
rolling down its potential \cite{Zlatev} appears to be
the most attractive. But most of these investigations are effective only for a
spatially flat ($k=0$) model and thus the fine tuning or the flatness
problem remains an unsolved one. The exception is the recent work by
Chimento {\it et al.} \cite{Chimento}, where a combination of a quintessence 
scalar field with a potential and a fluid with a bulk viscous stress 
has been shown to work for  $k = -1 $. This approach also solves 
the coincidence problem in the sense that the ratio of the density 
parameters for the normal matter and the scalar field asymptotically 
becomes a constant.

The aim of this paper is to show that Brans-Dicke's scalar tensor theory
can potentially solve the quintessence problem as it can lead to 
non-decelerating solutions for the scale factor for the present universe
dominated by cold matter with negligible pressure. We would like to
stress that this kind of solutions can be  obtained without
invoking a Q-matter or even dissipative processes. For a spatially
flat model ($k=0$) one can indeed have an accelerated expansion, and
even for a non-flat ($k \neq 0$) model, one can at least have a 
non-decelerated expansion. Very recently Bertolami and Martins \cite{Bertolami}
presented an accelerating model for the spatially flat $(k=0)$ in a
modified  Brans-Dicke theory using a potential which is a function of the 
Brans-Dicke scalar field itself. In this work we show that this 
solution also can be obtained in Brans-Dicke theory   without the 
potential.   However, there are problems in bridging this result
with the radiation-dominated decelerated universe for the same values
of the Brans--Dicke parameter $\omega$. In the next section we show that
the solution to this problem lies in the natural generalization of the
theory by allowing $\omega$ to be a function of the Brans--Dicke 
scalar field.

\section{Field equations and solutions}
For a Friedmann--Robertson--Walker spacetime, with scale factor $a(t)$,
spatial curvature index $k$, and assuming that the only matter--field 
is a perfect fluid, the gravitational field equations in Brans--Dicke 
theory are 
\be
3\frac{\dot{a}^2+k}{a^{2}} = {\rho_{f}\over\phi} 
-3{\dot {a} \dot{\phi}\over{a \phi}}
+{\omega\over{2}}{\dot{\phi}^{2}\over{\phi}^{2}},
\n{fe1}
\ee

\be
2{\ddot{a}\over{a}} + \frac{\dot{a}^{2}+k}{a^{2}} = {-p_{f}\over{\phi}} 
-{\omega\over{2}}{\dot{\phi}^{2}\over{\phi^{2}}} 
-2{\dot{a}\dot{\phi}\over{a\phi}} - {\ddot{\phi}\over{\phi}},
\n{fe2}
\ee
where $\omega$ is the Brans--Dicke's constant parameter, while $\rho_{f}$ 
and $p_{f}$ are the density and hydrostatic pressure respectively of the 
fluid distribution. These two latter quantities are connected by 
a barotropic equation of state $ p_{f} = (\gamma_{f} - 1)\rho_{f}$,
$\gamma_{f}$ being the (constant) adiabatic index of the fluid. 

The wave equation for the Brans--Dicke scalar field reads

\be
{\ddot{\phi}} + {3\dot{a}\dot{\phi}\over{a}}
=\frac{{\rho_{f}} - 3p_{f}}{2{\omega}+3},
\n{wave}
\ee
and this combined with the field equations  (~\ref{fe1}) and 
(~\ref{fe2}) leads to the matter conservation equation

\be
\dot{\rho}_{f} + 3{\dot{a}\over{a}}({\rho_{f}} + p_{f} ) = 0.
\n{mceq1}
\ee

Assuming that at the present epoch the universe is filled with cold
matter of negligible pressure (i.e., dust), we put $p_{f}=0$ in the 
last equation and obtain the relation
\be
\rho_{f}=\rho_{0}a^{-3},
\n{mceq2}
\ee
where $\rho_{0}$ is a constant of integration. With
$p_{f}=0$ and $\rho_{f}$ being given by equation (~\ref{mceq2}), the 
wave equation becomes
\[
{(a^{3}\dot{\phi})}^{.} = {{\rho_{0}}\over{2 \omega}+3}, 
\]
which immediately  yields the first integral
\be
a^{3}\dot{\phi} = {{{\rho_{0}}t}\over{2 \omega+3}}.
\n{fint}
\ee

We are primarily interested in a power law non--decelerating solution
for the scale factor in keeping with the recent observations. So
we take 
\be
a=a_{0}t^{\alpha}  \qquad \quad (\alpha \geq 1)  
\n{sf}
\ee
where $a_{0}$ is a positive--definite constant.

Equation (\ref{fint}) can now be integrated for $\phi$ as 
\be
\phi = \frac{\rho_{0} \, t^{2-3\alpha}}{a_{0}^{3}(2\omega +3)(2-3\alpha)}.
\n{sint}
\ee

Inspection of the field equations (~\ref{fe1}) and (~\ref{fe2})
reveals that when $k\neq{0}$, the only possible solution 
in this simple power function form is that with $ \alpha = 1$, 
i.e., 
\be
a = a_{0} t ,
\n{a1}
\ee
and

\be
\phi = -{{\rho}_{0}\over{a_{0}^{3} (2\omega + 3)t}}.
\n{phi1}
\ee
So one has a scenario where the present deceleration parameter
vanishes ($q_{0} = 0$), i.e., today the Universe is in a state of
uniform expansion. Although this very simple model does
not yield an accelerating universe for a non--zero $k$, it expands 
with no deceleration either and may 
sufficiently fulfil the requirements set by the recent 
observations on the distant supernovae \cite{Riess}. 

From (\ref{fe2}) the consistency condition (for $\alpha = 1$, i.e., 
$q_{0} = 0$)
\be
\omega = -2(1+\frac{k}{a_{0}^{2}})
\n{cc}
\ee
follows. The latter equation indicates that at least for $k=0$ or
$k=+1$ , $ \omega$ is negative. 
This is perfectly consistent with equation (~\ref{phi1}) which 
requires that $\omega < - 3/2$  as $\phi$ is a positive quantity.

If the universe is spatially flat $(k=0)$, this model yields several
possibilities including accelerated expansions for the universe. With
$k=0$, the equation (~\ref{fint}), put back in the field equations 
yields for dust ($p_{f}=0$) the general solution  as \cite{Brans} 
$\alpha = 2(\omega + 1)/(3 \omega + 4)$, i.e.,

\be
a = a_{0} \, t^{2({\omega}+1)/(3{\omega}+4)},
\n{a2}
\ee
and the scalar field takes the form

\be
\phi = \frac{(3\omega+4)\, \rho_{0} \, t^{2/(3\omega+4)}}{2 a_{0}^{3} 
(2\omega + 3)}.
\n{phi2}
\ee
Now it is easily seen that for different negative values of $\omega$,
one can generate different accelerating solutions. Such as, for
$\omega =-5/3$, we get $\alpha = 4/3$ and $\phi\propto t^{-2}$, which is
the solution presented by Bertolami and Martins \cite{Bertolami}. 
In this case, $q_{0} = -1/4$. For ${\omega} = -8/5$, $\alpha$ becomes $3/2$
and $\phi \propto t^{-5/2}$, giving $q_{0} = -1/3$, i.e., an accelaration 
rate higher than the Bertolami-Martins' solution.
For different values of $\omega$ in the range $-2 \leq{ \omega} \leq{-3/2}$, 
this model yields a host of accelerating solutions for a spatially flat 
universe. In fact this negative value of $\omega$ explains the behaviour 
of the model. A sufficiently negative $\omega$ may effectively lead to a
negative pressure (see equation (\ref{fe2})) and thus drive a positive 
acceleration or uniform expansion  without the violation of the
energy condition by normal matter.\\ \\
The solutions and the corresponding value of $\omega$ can be used in
equation (~\ref{fe1}) to calculate the age $t_{0}$ of the universe.
It turns out that the age is of the order of  $H_{0}^{-1}$. For the
uniformly expanding spatially flat universe, $t_{0}$ is exactly equal
to $H_{0}^{-1} \sim 1.62\times{10^{10}}$ yr. The higher the acceleration, 
the older the universe is. For $q_{0} = -1/4$, $t_{0}=\frac{4}{3}H_{0}^{-1}$.\\
As Brans--Dicke theory is a varying $G$  theory, the rate of this variation 
has been checked in this model and it appears to be
compatible with the observational limit. For instance, with $q_{0} = -1/4$,
$(\dot{G}/G)_{0}= 3H_{0}/2 < 2.5 \times 10^{-10}$ 
which is safely below the upper limit of 
$4 \times 10^{-10} \, $   yr$^{-1}$  set by observations 
(see \cite{Weinberg} and references therein). \\

It is important to note that although for the some negative values of
$\omega$, the model can lead to decelerating expansions for the
Universe when it is radiation--dominated, in that case the value 
of $\omega$ would be less negative than $-3/2$, i.e., 
$-3/2\leq{\omega}\leq{0}$. The indicated range of values of
$\omega$, i.e., $-2 \leq{ \omega} \leq{-3/2}$, which drives an
accelerated expansion for a matter--dominated model, does not produce
a consistent model with the radiation--dominated epoch. Thus the suggested 
model seems to badly spoil the big--bang nucleosynthesis scenario. One way 
out of this problem is to start with a modified version of Brans--Dicke 
theory where the parameter $\omega$ is a function of the scalar field $\phi$ 
rather than a constant \cite{Nordtvedt}. In this case equations (\ref{fe1}),
(\ref{fe2}) and (\ref{mceq1}) remain in place, but the wave equation for
the scalar field (\ref{wave}) has an additional term and becomes
\be
{\ddot{\phi}} + {3\dot{a}\dot{\phi}\over{a}}
=\frac{\rho_{f} - 3p_{f}}{2 \omega + 3} - \frac{\dot{\omega}\dot{\phi}}
{2\omega +3}.
\n{wave2}
\ee
\noindent
With the equation of state for radiation ($p_{f} = 
\textstyle{1\over{3}} \rho_{f}$), equation (\ref{wave2}) immediately
gives a first integral
\be
\dot{\phi} (2\omega + 3)^{1/2} = \frac{A}{a^{3}},
\n{fint1}
\ee
\noindent
where $A$ is an integration constant.
It turns out that there are functional forms of $\omega$ for which 
the universe expands with a deceleration such as like $a \propto 
{t^{\frac{1}{2}}}$ in this case so that the primordial
nucleosynthesis can be successfully explained.
A simple choice like
\be
2\omega + 3 = (\phi - 1)^{2}
\n{simple}
\ee
\noindent
when used in equation (\ref{fint1}) along with $a \propto t^{1/2}$ yields
the integral
\be
(\phi - 1)^{2} = \frac{B}{\sqrt{t}},
\n{yields}
\ee
\noindent
$B$ being a constant of integration. Equations (\ref{simple}) and 
(\ref{yields}) clearly shows that $\omega$ decreases with time
towards a constant value $ -3/2$, which indeed produces an accelerated
expansion for a late dust--dominated universe as we have seen. \\  \\
In general, a choice of $\omega$ as a polynomial in $\phi$ is
\be
2\omega +3 = \sum_{i=1}^{n} A_{i} \, \phi^{n_{i}}, 
\n{omega}
\ee
\noindent
where $n_{i} \geq 0$ and $A_{i}$'s are constants appears to solve
this problem. A varying 
$\omega$ theory with a nonminimal coupling between gravity and the scalar 
field different from the Brans--Dicke theory ($\phi^{2} R$ as opposed to 
$\phi R$  in BDT) may also explain the late time behaviour of the universe 
as pointed out very recently by Bartolo and Pietroni\cite{Bartolo}.
It is also worthwhile to note that Sen and Seshadri\cite{sen} have shown 
that with small negative values of $\omega$, it is possible to obtain
growing modes for the density perturbations for a universe with a power
law expansion rate.

\section{A conformally transformed version}
In Brans-Dicke theory, the folklore is that $\omega$ is positive, or
at least larger than $-3/2$. There are two reasons for this belief.
The first reason is that in the weak field limit, the Newtonian
constant of gravitation G is given by,
\[
G = \left(\frac{2\omega+3}{2\omega+4}\right) \frac{1}{\phi},
\]
and for $-2< \omega <-3/2$, G becomes negative. But it should be
emphasized that this is only in a weak field limit. So in the model
presented in the present work, G does not become negative, only this
weak field approximation does not hold. In the full non linear
Brans-Dicke theory, the effective Gravitational constant is actually
given by
\[
G = \frac{G_{0}}{\phi},
\]
which is indeed positive as $\phi$ is positive. It deserves mention that for 
the accelerating solution obtained by Bertolami and Martins \cite{Bertolami}
also, $\omega$ is actually negative.

The second reason for the belief that $\omega > -3/2$ is that the
energy related to this scalar field is proportional to
$(2 \omega+3)$. To see this, it is better to effect a 
conformal transformation

\[
{\bar{g}}_{\mu \nu} = {\phi}g_{\mu \nu}.
\]
In the transformed version, G becomes a constant, but one has to
sacrifice the equivalence principle as the rest mass of a test
particle becomes a function of the scalar field \cite{Dicke}. 
So the geodesic equations are no longer valid 
and naturally the physical significance of
different quantites are indeed questionable in this version of the
theory. Nevertheless the equations in this version gives an insight in
comparing the energies of different components of matter. The
equations (~\ref{fe1})--(~\ref{fe2}) in the new frame look 
like

\be
{3(\dot{\bar{a}}^{2} + k)\over{\bar{a}^{2}}} = \bar{\rho}_{f} + 
{\frac{2{\omega}+3}{4}}{\dot{\psi}^{2}} ,
\n{nf1}
\ee

\be
{{2\ddot{\bar{a}}}\over{\bar{a}}}+{{{\dot{\bar{a}}^{2}}+k}\over{\bar{a}^{2}}}
= -\bar{p}_{f} -{\frac{2{\omega}+3}{4}}{\dot{\psi}^{2}} ,
\n{nf2}
\ee
and the wave equation (~\ref{wave}) transforms to 

\be
\ddot{\psi} + 3 \frac{\dot{\bar{a}}\dot{\psi}}{\bar{a}} =
\frac{\bar{\rho}_{f} -3\bar{p}_{f}}{2\omega + 3},
\n{nf3}
\ee
where an overhead bar indicates quantities in the new frame,
$\psi = \mbox{ln} \phi$, and  $\bar{a}^{2}=\phi a^{2}$. The density and 
the pressure of the normal matter in this version
are related to those in the original version in a simple way as 
$\bar{\rho}_{f}=\phi^{-2} \rho_{f}$ and $\bar{p}_{f}=\phi^{-2}p_{f}$. It is 
clearly seen that the contribution to the energy density by the scalar 
field is given by 

\be
\bar{\rho}_{\phi} = \frac{2\omega + 3}{4} {\dot{\psi}}^{2}.
\n{rphi}
\ee
Here $\psi$ looks like a massless minimally coupled scalar field and
at least formally behaves as a perfect fluid with a stiff equation state,
$\bar{p}_{\phi} = \bar{\rho}_{\phi}$. (Note that the corresponding adiabatic 
index is $\gamma_{\phi} = 2$).
Equation (~\ref{rphi}) explains why $\omega$ is usually taken to be 
larger than the $-3/2$ from the consideration of the positivity of 
energy. 
 
Our intention is to get a sufficiently negative pressure from some
source so that we can get $q_{0}\leq 0$, and here we achieve that by
means of a scalar field $\phi$ which has a negative energy. But in our
case $\phi$ is a geometric field unlike the non--gravitational fields
such as normal matter or the quintessence field, and so its having a
negative energy is not pathological in that sense. We rather save the
normal matter from the unappealing feature of a negative pressure and
yet get a non positive--definite deceleration parameter. 

This conformally transformed version of the theory gives an insight regarding 
the solution of the flatness problem too. The density parameter is defined as
$\bar{\Omega} = \bar{\rho}/3\bar{H}^{2}$, where 
\[
\bar{\rho} \equiv \bar{\rho}_{f} + \bar{\rho}_{\phi} =\bar{\rho}_{f}  
+ \frac{2\omega + 3}{4} {\dot{\psi}}^{2},
\] 
is the total energy density.
The subscripts $f$ and $\phi$ refer to the normal fluid and the
scalar field, respectively. $\bar{H} \equiv {\dot{\bar{a}}/{\bar{a}}}$
is the Hubble parameter in the new  frame. 

The Bianchi identity yields the relation

\be
\dot{\bar{\rho}} +3{\gamma} \bar{H} \bar{\rho} = 0,
\n{brho}
\ee
where $\gamma$ is the average barotropic index defined by the relation
\be
{\gamma}\bar{\Omega}={\gamma}_{f}{\bar{\Omega}}_{f} + 
{\gamma}_{\phi}\bar{\Omega}_{\phi} ,
\n{average}
\ee
where $\bar{\Omega} \equiv \bar{\Omega}_{f} + \bar{\Omega}_{\phi}$.

Using equations (~\ref{nf1}) and (~\ref{brho}) one can write
\be
\dot{\bar{\Omega}} = {\bar{\Omega}}({\bar{\Omega}}-1)(3{\gamma}-2)\bar{H}.
\n{bomega}
\ee
This equation shows that $\bar{\Omega} = 1$ is indeed a solution, but 
$ \left(\partial \dot{\bar{\Omega}}/\partial{\bar{\Omega}}\right)_{H}$
must be negative at $\bar{\Omega} = 1$ if this $\bar{\Omega}$ value is to 
be a stable solution of (~\ref{bomega}). 
This requires that $\gamma$ should be less than $2/3$. 
As the ratio of the density to pressure of the fluid remains the same in the 
two versions of the theory, the adiabatic  index $\gamma_{f}$ also remains 
the same. Because of our choice of $\bar{p}_{f} =0$ and 
${\bar{p}}_{\phi} = {\bar{\rho}}_{\phi}$, we have 
${\gamma}_{f} =1$ and ${\gamma}_{\phi} = 2$. So from equation
(~\ref{average}) one has 

\be
\gamma = \frac{\bar{\Omega}_{f} +
2\bar{\Omega}_{\phi}}{\bar{\Omega}_{f} + \bar{\Omega}_{\phi}}.
\n{gamma}
\ee 
As $2\omega +3$ is negative, $\bar{\Omega}_{\phi}$ is also negative
and one can achieve $\gamma < 2/3 $ provided

\[
\bar{\Omega}_{f} < 4 \mid \bar{\Omega}_{\phi}\mid.
\]
So $\bar{\Omega} = 1$ is indeed a stable solution in this model
depending on the relative magnitudes of the energies of matter and
the Brans-Dicke's scalar field. With $\bar{\Omega} = 1$, we have 
$\bar{\Omega}_{k} =- k/\bar{a}^{2} = 0$ and thus the flatness problem 
can be solved.

It is true that although the conformally transformed version 
(popularly known as the Einstein frame) is appealing from the point
of view of the computational simplicity, the quantities lose their
physical significance for reasons stated at the beginning of this section. 
But it must be emphasized that the character of $k$
remains the same and if it is zero in this version, it is so in the
physical original version of the theory as well.
For a spatially flat universe ($k=0$) we can thus construct
accelerating models for the Universe and reproduce the
Bertolami-Martins solution as a special case and we do not have
to invoke an additional self--interaction term for this. Even for 
the non-flat cases, we can at least obtain a non--decelerating 
model with $q_{0} = 0$, whereas almost all the quintessence models
produce solutions only for a spatially flat universe with the only
exception of the dissipative fluid model recently discussed by 
Chimento {\it et al.} \cite{Chimento}, which works for open 
universes as well. For a discussion of the flatness problem in
Brans--Dicke theory see the paper by Levine and Freese \cite{levin}.

\section{Concluding remarks}
Brans--Dicke theory proved to be extremely useful in solving some of
the outstanding problems in the inflationary universe scenario with
the possibility of an ``extended inflation" \cite{Mathiazhagan}. 
And now although the theory  might relinquish some of its natural 
attraction as it fails to produce general relativity as an 
infinite $\omega$ limit \cite{Banerjee} contrary to the earlier 
belief, once again it appears to be able to account for some outstanding
features of our present Universe like the accelerated expansion and 
the coincidence problem by its own right without having to invoke 
dissipative processes or exotic fields.

It should not be overlooked, however, that $\omega$ has to pick up a low 
negative value (i.e., $- \omega \sim {\cal O}(1)$) in order to solve the 
acceleration and coincidence problems by a single stroke (e.g., 
$q_{0} = - 1/4$ implies $a \propto t^{4/3}$ and consequently $\omega = -5/3$). 
This squarely conflicts with the lower limit imposed on $|\omega|$ by 
solar system experiments, namely $|\omega| \geq 500$ \cite{Will}. 
Nevertheless in ``extended inflation" the model of La and Steinhardt
worked provided that $\omega$ takes a value close to $20$ 
\cite{Mathiazhagan} which is also embarrasingly lower than the 
mentioned astronomical limit. Therefore it remains a problem
to find a suitable compromise between astronomical observations and 
cosmological requirements. Another problem is that the constant negative
value of $\omega$ does not produce a consistent radiation model which
explains the primordial nucleosynthesis. We have shown that a varying
$\omega$ theory like Nordtverdt's \cite{Nordtvedt} can give rise to a
decelerating radiation model where the big--bang nucleosynthesis 
scenario is not adversedly affected and $\omega$ evolves to the small
and negative constant values required for the late time acceleration
in the matter--dominated epoch. A thorough survey of varying $\omega$
theories is perhaps warranted which may bridge the decelerating radiation 
universe with the accelerating matter universe while the local 
inhomogeneities might locally give rise to high values of $\omega$ to 
be consistent with astronomical experiments.

However, attempts to solve the aforesaid problems outside
Brans--Dicke theory do not fare much better. The introduction of
a cosmological constant of the order of the critical density lacks 
of a solid support from quantum field theory \cite{Weinberg1}, and 
arguments based on the anthropic principle are anything but 
convincing \cite{Weinberg2}. On their part models based on 
quintessence suffer from the problem of unwanted long--range forces
and that they cannot be as homogeneous as it should \cite{Carroll}.

\acknowledgments 
The authors thank Winfried Zimdahl for a critical reading of an earlier
version of this paper and valuable comments. This work has been partially 
supported by Spanish Ministry of Education under Grant PB94--0781. One 
of us (NB) is grateful to the ``Direcci\'o General de la Recerca" of 
the Catalonian Government for financial support under grant PIV99.

\end{document}